\documentclass[english,aps]{revtex4}
\usepackage[T1]{fontenc}
\usepackage[latin9]{inputenc}
\usepackage{array}
\usepackage{graphicx}
\usepackage{esint}

\makeatletter

\providecommand{\tabularnewline}{\\}

\include{graphics}

\makeatother

\usepackage{babel}

\begin{document}

\preprint{Alberta Thy 09-08}

\title{Semi-Leptonic $b$-decay at Intermediate Recoil}

\author{Matthew Dowling}
\author{Alexey Pak}
\author{Andrzej Czarnecki}
\affiliation{Department of Physics, University of Alberta, Edmonton, Alberta,
Canada T6G 2G7}

\begin{abstract}
We compute the ${\mathcal{O}}\left({\alpha_s^2}\right)$ corrections to the differential
rate of the semileptonic decay $b\to c\ell \nu_\ell$ at the
``intermediate recoil'' point, where the $c$-quark mass and the
invariant mass of the leptons are equal.  The calculation is based on
an expansion around two opposite limits of the quark masses $m_{b,c}$:
$m_c\simeq m_b$ and $m_c\ll m_b$.
The former case was previously studied; we correct and extend that
result.  The latter case is  new.  The smooth matching of both
expansions provides a check of both.  We clarify the discrepancy
between the recent determinations of the full NNLO QCD correction to
the semileptonic $b\to c$ rate, and its earlier estimate. 

\end{abstract}

\maketitle

\section{Introduction}

Approximately one out of five decays of the $b$-quark produces a
$c$-quark accompanied by leptons. Those semileptonic decays provide
information about quark masses, the Cabibbo-Kobayashi-Maskawa (CKM)
matrix element $V_{cb}$, as well as properties of hadrons containing
heavy quarks. In order to access that information, measurements of
the decay probability and distributions are compared with theoretical
predictions that account for radiative corrections, quark masses,
and non-perturbative effects of strong interactions. Given that the
strong coupling constant at the mass-scale of the $b$-quark is quite
large, $\alpha_{s}\left(q^{2}=m_{b}^{2}\right)\simeq0.2$, and that
the present uncertainty in $V_{cb}$ approaches the one-percent level
\cite{Amsler:2008zz}, it is important to determine the second-order
effects, $\mathcal{O}\left(\alpha_{s}^{2}\right)$. 

The full $\mathcal{O}\left(\alpha_{s}^{2}\right)$ correction to the
decay rate was first calculated in the limit of a massless produced
quark (relevant for the decay $b\to u\ell\nu_{\ell}$)
\cite{vanRitbergen:1999gs}.  Effects of the $c$-quark mass were known,
until recently, only in the so called Brodsky-Lepage-Mackenzie (BLM)
approximation \cite{Brodsky:1982gc,Luke:1994yc}, estimating the
largest part of the second order corrections using the running of
$\alpha_{s}$. The remaining, non-BLM corrections, are usually smaller
and much more difficult to determine.  They were known only in three
special points of lepton kinematics: the zero recoil, where the
leptons are emitted back-to-back and the produced quark remains at
rest \cite{Czarnecki:1997cf,Czarnecki:1996gu,Franzkowski:1997vg}; the maximum recoil, with
the vanishing invariant mass of the leptons
\cite{Czarnecki:1997hc,Pak:2006xf}; and the intermediate recoil, where
the invariant mass of the leptons equals that of the $c$-quark
\cite{Czarnecki:1998kt}. In the latter study, the information from all
three points was used to estimate the
$\mathcal{O}\left(\alpha_{s}^{2}\right)$ correction to the total decay
rate with a massive $c$-quark.

Very recently, two independent studies determined the full mass dependence
of the non-BLM corrections: in \cite{Melnikov:2008qs}, the calculation
was performed numerically for arbitrary quark masses, and in \cite{Pak:2008qt}
an expansion around small $m_{c}/m_{b}$ was obtained analytically.
The two methods are very different, with the former being more accurate
at large, and the latter at small $m_{c}$, but they agree very well
in the physically interesting region of $m_{c}=\left(0.25\ldots0.30\right)m_{b}$.
Unfortunately, the resulting non-BLM correction disagrees almost by
a factor of two with the estimate found in \cite{Czarnecki:1998kt}. 

\begin{figure}[!b]
\centering\includegraphics[width=8cm]{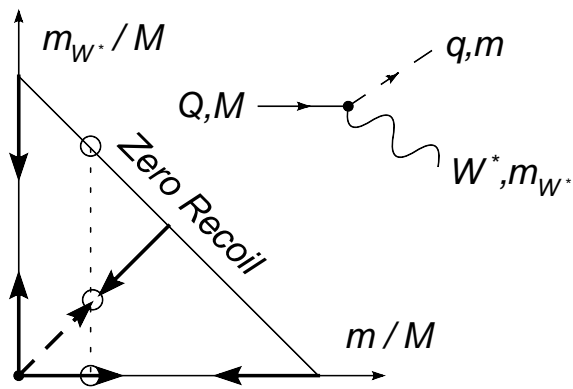}\caption{\label{fig:Triangle}The kinematic region where the decay $Q\to qW^{*}$
is allowed. The solid arrows show known expansions while the dashed
arrow shows the expansion presented here. The decay width is also
known analytically along the whole zero recoil line. The dotted line
corresponds to the decay width $Q\to q\ell\nu_{\ell}$. The three
circles along this line show known values coming from the different
expansions. In the case considered in this paper, $M=m_b$ and $m=m_c$.}

\end{figure}

The goal of the present paper is twofold. First, we want to check the
intermediate-recoil expansion presented in \cite{Czarnecki:1998kt}.
Among the three kinematical points on which the estimate
\cite{Czarnecki:1998kt} of the total correction was based, the
intermediate-recoil is the only one not checked by an independent
calculation. An expansion is constructed from the opposite limit than
in \cite{Czarnecki:1998kt}: whereas there the expansion was around the
zero-recoil limit, here we start from the vanishing $m_{c}$, as shown
in Fig. \ref{fig:Triangle}.  In addition, the old expansion around the
zero-recoil limit is repeated and extended to higher orders. Our
second goal is to clarify the source of the disagreement between the
three-point estimate and the recent explicit calculations.

Fig.~\ref{fig:Triangle} puts the present expansion in the context of
the possible kinematics of a heavy to light quark decay, $Q\to q +
\left(W^{*}\to\ell\nu_{\ell}\right)$. Along the diagonal, the mass of
the virtual $W^{*}$ is equal to the light-quark mass. The arrow
originating from the zero-recoil line corresponds to the expansion
done in \cite{Czarnecki:1998kt} (and repeated in the present paper),
while the dashed arrow coming from the zero mass point corresponds to
the expansion presented here for the first time. Ultimately, these
expansions should give a consistent value for the decay width
$\Gamma(b\to cW^{*})$. It is related to the differential semi-leptonic
width, \begin{equation} \frac{d\Gamma(b\to
c\ell\nu_{\ell})}{dq^{2}}=\frac{G_{F}}{6\pi^{2}\sqrt{2}M_{W}^{2}}q^{2}\Gamma(b\to
cW^{*})|_{m^{2}(W^{*})=q^{2}},\label{eq:Integral}\end{equation} where
$q^{2}$ is the invariant mass squared of the leptons, and Fermi
constant is $G_{F}=\frac{\sqrt{2}g_{w}^{2}}{8M_{W}^{2}}$.

\section{Expansion From Zero Mass Point
\label{sec:Expansion-From-Zero}}

Using the intermediate-recoil relation $m_{W^{*}}=m_{c}$, we here
calculate the width as a series in $\rho\equiv\frac{m_{c}}{m_{b}}\ll1$,
and $\alpha_{s}$,
\begin{eqnarray}
\Gamma(b\to cW^{*}) & = & \Gamma_{0}\left[X_{0}+C_{F}\frac{\alpha_{s}}{\pi}X_{1}
+C_{F}\left(\frac{\alpha_{s}}{\pi}\right)^{2}X_{2}+\mathcal{O}\left(\alpha_{s}^{3}\right)\right],
\label{eq:widthIntoWstar}
\end{eqnarray}
where
\begin{equation}
\Gamma_{0}=\frac{g_{w}^{2}|V_{cb}|^{2}m_b^{3}}{64\pi m^{2}(W^{*})}.
\end{equation}
The tree-level and first-order results $X_{0,1}$ are known exactly
\cite{Jezabek:1988iv,Czarnecki:1990kv}, and the present approach,
described below, has been tested with them up to $\mathcal{O}(\rho^{10})$,\begin{eqnarray}
X_{0} & = & \left(1-\rho^{2}\right)\sqrt{1-4\rho^{2}},\\
X_{1} & = & \frac{5}{4}-\frac{\pi^{2}}{3}+\rho^{2}\left(\frac{\pi^{2}}{3}-\frac{5}{4}-9\ln\rho\right)+\rho^{4}\left(9\ln\rho-\frac{15}{4}\right)+\ldots.\end{eqnarray}

\begin{figure*}[t]
\centering\includegraphics[width=0.3\textwidth]{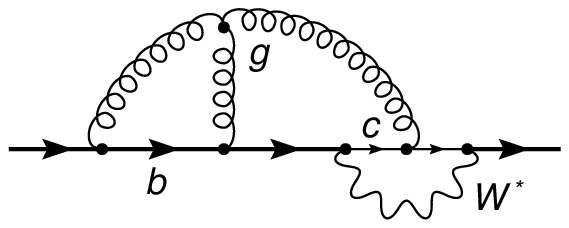} \includegraphics[width=0.3\textwidth]{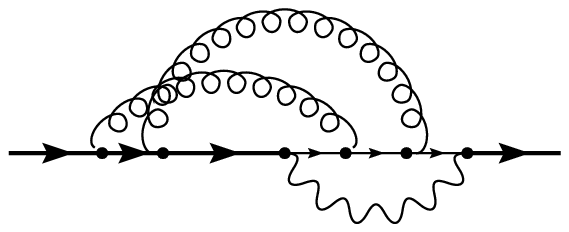}
\includegraphics[width=0.3\textwidth]{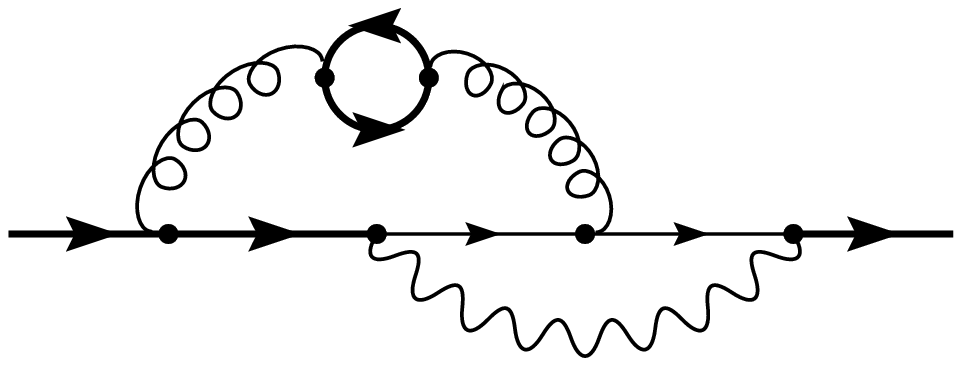}

\caption{\label{fig:Diagrams}A sample of the diagrams needed
 for the intermediate-recoil expansion presented here.}

\end{figure*}

To evaluate the $\mathcal{O}(\alpha_{s}^{2})$ corrections, we considered
the imaginary parts of 39 three-loop self-energy diagrams with massive
propagators, such as in Fig.~\ref{fig:Diagrams}, and used the optical
theorem to calculate the decay width.%
To deal with the two scales, $m_{b}$ and $m_{c}$, we used the method
of asymptotic expansion \cite{Tkachov:1997gz,Smirnov:2002pj}. As
an example of how this asymptotic expansion is done, consider the
left hand diagram in Fig.~\ref{fig:Diagrams}. %
\begin{figure*}
\begin{tabular}{|>{\raggedright}p{3cm}|>{\raggedright}p{5cm}|>{\raggedright}p{6cm}|}
\hline 
\parbox[t][1\totalheight]{3cm}{%
Topology%
}

\includegraphics[width=2.9cm]{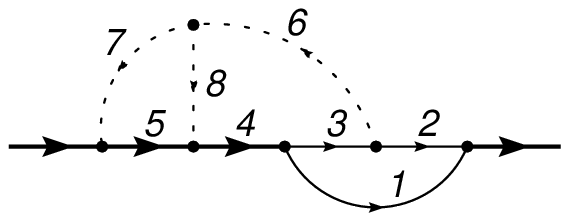} & \multicolumn{2}{l|}{%
\parbox[t][1\totalheight]{11cm}{%
$[1]=k_{1}^{2}+m_c^{2}$ $[2]=k_{2}^{2}+m_c^{2}$ $[3]=k_{3}^{2}+m_c^{2}$ 

$[4]=(k_{3}-k_{1})^{2}+m_b^{2}$ $[5]=k_{4}^{2}+m_b^{2}$ $[6]=(k_{3}-k_{2})^{2}$ 

$[7]=(k_{4}-p)^{2}$ $[8]=(p+k_{3}-k_{2}-k_{4})^{2}$\smallskip{}
}}\tabularnewline
\hline 
\parbox[t][1\totalheight]{3cm}{%
 Region 1 %
}

\includegraphics[width=2.9cm]{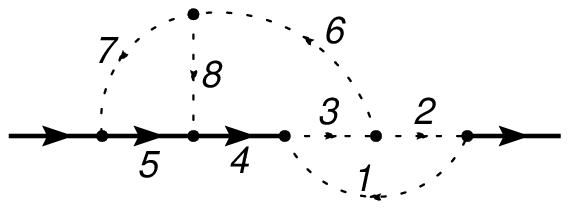} & %
\parbox[t][1\totalheight]{5cm}{%
$|k_{1}|\gg m_c$, $|k_{2}|\gg m_c$, $|k_{3}|\gg m_c$%
} & %
\parbox[t][1\totalheight]{6cm}{%
$[1]\to k_{1}^{2}$, $[2]\to k_{2}^{2}$, $[3]\to k_{3}^{2}$%
}\tabularnewline
\hline 
\parbox[t][1\totalheight]{3cm}{%
 Region 2 %
}

\includegraphics[width=2.9cm]{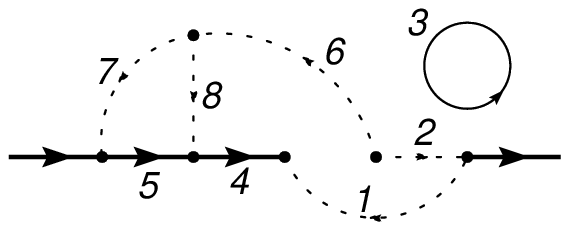} & %
\parbox[t][1\totalheight]{5cm}{%
$|k_{1}|\gg m_c$, $|k_{2}|\gg m_c$, $|k_{3}|\sim m_c$%
} & %
\parbox[t][1\totalheight]{6cm}{%
$[1]\to k_{1}^{2}$, $[2]\to k_{2}^{2}$, $[4]\to k_{1}^{2}+m_b^{2}$

$[6]\to k_{2}^{2}=[2]$, $[8]\to(p-k_{2}-k_{4})^{2}$%
}\tabularnewline
\hline 
\parbox[t][1\totalheight]{3cm}{%
 Region 3 %
}

\includegraphics[width=2.9cm]{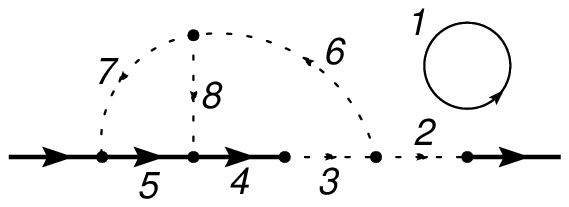} & %
\parbox[t][1\totalheight]{5cm}{%
$|k_{1}|\sim m_c$, $|k_{2}|\gg m_c$, $|k_{3}|\gg m_c$%
} & %
\parbox[t][1\totalheight]{6cm}{%
$[2]\to k_{2}^{2}$, $[3]\to k_{3}^{2}$, $[4]\to k_{3}^{2}+m_b^{2}$%
}\tabularnewline
\hline 
\parbox[t][1\totalheight]{3cm}{%
 Region 4 %
}

\includegraphics[width=2.9cm]{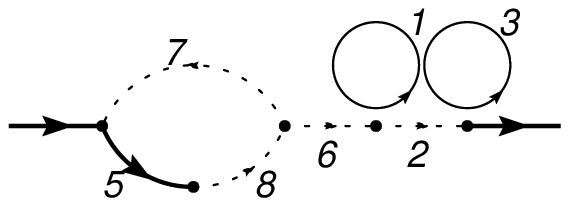} & %
\parbox[t][1\totalheight]{5cm}{%
$|k_{1}|\sim m_c$, $|k_{2}|\gg m_c$, $|k_{3}|\sim m_c$%
} & %
\parbox[t][1\totalheight]{6cm}{%
$[2]\to k_{2}^{2}$, $[4]\to m_b^{2}$, $[6]\to k_{2}^{2}$,

$[8]\to k_{4}^{2}$%
}\tabularnewline
\hline
\end{tabular}

\caption{\label{fig:Asymptotic}The asymptotic expansion of a two-scale diagram
used to integrate the left-hand diagram in Fig.~\ref{fig:Diagrams}.
The dashed, thin and thick lines correspond to massless, soft-scale
massive and hard-scale massive propagators respectively. }

\end{figure*}
 We consider {}``regions'' where each loop momentum is either hard,
$\sim m_{b}$, or soft, $\sim m_{c}$, and Taylor expand the propagators
so that in the end we only have to deal with single scale diagrams
as shown in Fig.~\ref{fig:Asymptotic}. This method produced as many
as 11 regions for a single topology. Expansions to the desired order
$\mathcal{O}(\rho^{10})$ required algorithm \cite{Laporta:2001dd}
for the un-factorized three loop regions, (e.g. Region 1 in Fig.~\ref{fig:Asymptotic}).

The second order results can be separated into a sum of gauge invariant
parts, each with a different color factor,
\begin{eqnarray}
X_{2} & = & T_{R}(N_{L}X_{L}+N_{S}X_{S}+N_{H}X_{H})
+C_{F}X_{F}+C_{A}X_{A}.
\label{eq:X2colorParts}\end{eqnarray}
In this expression, the $X_{i}$'s are the gauge invariant parts in
terms of $\rho$, $N_{L}$ is the number of quarks lighter than a
$c$-quark, $N_{S}$ and $N_{H}$ serve as markers to separate the
$c$-quark and $b$-quark loop contributions. $C_{F}=\frac{4}{3}$,
$C_{A}=3$, and $T_{R}=\frac{1}{2}$, are the appropriate color factors
in $SU(3)$. The contributions from a top quark loop are not considered
here because they are suppressed by the ratio $m_{b}/m_{t}$ and are
negligible. Terms up to $\mathcal{O}(\rho^{10})$ have been calculated completely analytically.
Here we present the formulas for terms up to $\rho^{4}$
to save space, Eqs.~(\ref{eq:XH},~\ref{eq:XS},~\ref{eq:XL},~\ref{eq:XA},~\ref{eq:XF}),
while the numerical coefficients of all terms are given in Table~\ref{tab:rho}.

\begin{eqnarray}
X_{H} & = & \frac{12991}{1296}-\frac{\zeta_{3}}{3}-\frac{53\pi^{2}}{54}+\rho^{2}\left[\frac{89\pi^{2}}{54}-\frac{137567}{6480}+\frac{13\zeta_{3}}{3}\right]+\rho^{4}\left[\frac{4\pi^{2}}{3}-\frac{10081601}{705600}-\frac{23}{840}\ln\rho\right],\label{eq:XH}\\
X_{S} & = & \zeta_{3}-\frac{4}{9}+\frac{23\pi^{2}}{108}-\rho\frac{3\pi^{2}}{4}+\rho^{2}\left[\frac{4}{9}+\frac{13}{2}\ln\rho-3\ln^{2}\rho-\zeta_{3}+\frac{157\pi^{2}}{108}\right]-\rho^{3}\frac{25\pi^{2}}{18}\label{eq:XS}\\
 & + & \rho^{4}\left[\frac{1193}{36}-\frac{61}{3}\ln\rho+9\ln^{2}\rho-\frac{16\pi^{2}}{3}\right],\nonumber \\
X_{L} & = & \zeta_{3}-\frac{4}{9}+\frac{23\pi^{2}}{108}+\rho^{2}\left[\frac{13}{2}\ln\rho-\frac{1}{18}-3\ln^{2}\rho-\zeta_{3}-\frac{77\pi^{2}}{108}\right]+\rho^{4}\left[\frac{865}{72}-\frac{34}{3}\ln\rho+6\ln^{2}\rho\right],\label{eq:XL}\\
X_{A} & = & \frac{521}{576}+\frac{9\zeta_{3}}{16}+\frac{505\pi^{2}}{864}-\frac{19\pi^{2}}{8}\ln2+\frac{11\pi^{4}}{1440}-\rho^{2}\left[\frac{1223}{576}+\frac{185}{8}\ln\rho-\frac{33}{4}\ln^{2}\rho+\frac{107\zeta_{3}}{16}+\frac{145\pi^{2}}{864}\right.\label{eq:XA}\\
 & - & \left. \frac{57\pi^{2}}{8}\ln2+\frac{161\pi^{4}}{720}\right]+\rho^{3}\frac{2\pi^{2}}{3}+\rho^{4}\left[\ln\rho\left(\frac{2027}{48}-\frac{23\pi^{2}}{8}\right)-\frac{13391}{288}-\frac{33}{2}\ln^{2}\rho-\frac{403\zeta_{3}}{64}\right.\nonumber \\
 & + & \frac{27\pi^{2}}{8}-\left.\frac{201\pi^{2}}{32}\ln2-\frac{31\pi^{4}}{720}\right],\nonumber \\
X_{F} & = & 5-\frac{53\zeta_{3}}{8}-\frac{119\pi^{2}}{48}+\frac{19\pi^{2}}{4}\ln2-\frac{11\pi^{4}}{720}+\rho^{2}\left[\frac{151\zeta_{3}}{8}+\frac{743\pi^{2}}{48}+\ln\rho\left(\pi^{2}-\frac{75}{8}\right)-\frac{27}{2}\ln^{2}\rho\right.\label{eq:XF}\\
 & - & \left.\frac{97}{2}-\frac{57\pi^{2}}{4}\ln2-\frac{127\pi^{4}}{360}\right]-\rho^{3}\frac{4\pi^{2}}{3}+\rho^{4}\left[\frac{7145}{288}+\ln\rho\right(\frac{25\pi^{2}}{12}-\frac{329}{24}\left)+18\ln^{2}\rho+\frac{547\zeta_{3}}{32}\right.\nonumber \\
 & - & \left.\frac{83\pi^{2}}{12}+\frac{201\pi^{2}}{16}\ln2+\frac{19\pi^{4}}{72}\right].\nonumber \end{eqnarray}
\begin{table*}
\caption{\label{tab:rho}Numerical coefficients of the expansion presented
here to all orders calculated.}

\begin{tabular}{c>{\centering}p{1.2cm}>{\centering}p{1.2cm}>{\centering}p{1.2cm}>{\centering}p{1.5cm}>{\centering}p{1.2cm}>{\centering}p{1.3cm}>{\centering}p{1.5cm}>{\centering}p{1.5cm}>{\centering}p{1.2cm}>{\centering}p{1.5cm}>{\centering}p{1.5cm}}
\hline 
\noalign{\vskip0.5mm}
 & $\rho^{0}$ & $\rho^{1}$ & $\rho^{2}$ & $\rho^{2}\ln\rho$ & $\rho^{2}\ln^{2}\rho$ & $\rho^{3}$ & $\rho^{4}$ & $\rho^{4}\ln\rho$ & $\rho^{4}\ln^{2}\rho$ & $\rho^{5}$ & $\rho^{6}$\tabularnewline
\hline
\hline 
\noalign{\vskip0.5mm}
$X_{A}$ & $-8.154$ & - & $15.14$ & $-23.12$ & $8.25$ & $6.580$ & $-67.92$ & $13.85$ & $-16.5$ & $77.64$ & $-124.2$\tabularnewline
\noalign{\vskip0.5mm}
$X_{F}$ & $3.575$ & - & $-4.887$ & $0.4946$ & $-13.5$ & $-13.16$ & $88.74$ & $6.853$ & $18$ & $-155.3$ & $262.4$\tabularnewline
\noalign{\vskip0.5mm}
$X_{L}$ & $2.859$ & - & $-8.294$ & $6.5$ & $-3$ & - & $12.01$ & $-11.33$ & $6$ & - & $-12.45$\tabularnewline
\noalign{\vskip0.5mm}
$X_{S}$ & $2.859$ & $-7.402$ & $13.59$ & $6.5$ & $-3$ & $-13.71$ & $-19.50$ & $-20.33$ & $9$ & $30.98$ & $-13.50$\tabularnewline
\noalign{\vskip0.5mm}
$X_{H}$ & $-0.06360$ & - & $0.2460$ & - & - & - & $-1.129$ & $-0.02738$ & - & - & $1.656$\tabularnewline
\hline
\end{tabular}

\smallskip{}

\begin{tabular}{c>{\centering}p{1.5cm}>{\centering}p{1.5cm}>{\centering}p{1.5cm}>{\centering}p{1.5cm}>{\centering}p{1.5cm}>{\centering}p{1.5cm}>{\centering}p{1.5cm}>{\centering}p{1.5cm}>{\centering}p{1.5cm}>{\centering}p{1.5cm}}
\hline 
\noalign{\vskip0.5mm}
 & $\rho^{6}\ln\rho$ & $\rho^{6}\ln^{2}\rho$ & $\rho^{7}$ & $\rho^{8}$ & $\rho^{8}\ln\rho$ & $\rho^{8}\ln^{2}\rho$ & $\rho^{9}$ & $\rho^{10}$ & $\rho^{10}\ln\rho$ & $\rho^{10}\ln^{2}\rho$\tabularnewline
\hline
\hline 
\noalign{\vskip0.5mm}
$X_{A}$ & $-96.19$ & $14.17$ & $270.6$ & $-666.7$ & $-235.6$ & $40.01$ & $973.0$ & $-2327.3$ & $-705.7$ & $48.98$\tabularnewline
\noalign{\vskip0.5mm}
$X_{F}$ & $38.03$ & $-66.89$ & $-541.3$ & $1127.6$ & $-41.98$ & $-245.5$ & $-1945.9$ & $3771.9$ & $-516.1$ & $-733.0$\tabularnewline
\noalign{\vskip0.5mm}
$X_{L}$ & $19.30$ & $-6$ & - & $18.39$ & $35.59$ & $-18$ & - & $80.68$ & $101.0$ & $-76$\tabularnewline
\noalign{\vskip0.5mm}
$X_{S}$ & $15.77$ & $-12$ & $64.15$ & $-33.67$ & $44.76$ & $8$ & - & $-0.5973$ & $151.8$ & $34$\tabularnewline
\noalign{\vskip0.5mm}
$X_{H}$ & $-0.8866$ & - & - & $1.984$ & $-0.2800$ & - & - & $4.494$ & $0.5912$ & -\tabularnewline
\hline
\end{tabular}
\end{table*}
 For this expansion, we have used the $\overline{\mathrm{MS}}$ definition
of $\alpha_{s}$ normalized at the pole mass $m_{b}$.

\section{Expansion From the Zero-recoil Line}
\label{sec:zero}
An alternative way to compute at the intermediate recoil is to expand
around the zero-recoil limit where $m_{c}=m_{W^{*}}=\frac{m_{b}}{2}$.
The decay width parameterization in Eqs. (\ref{eq:Integral},\ref{eq:widthIntoWstar}),
as well as the decomposition of the second order correction into color
parts, Eq. (\ref{eq:X2colorParts}), are still valid. For the purpose
of the expansion around the zero-recoil limit, it is convenient to
parameterize the dependence on the quark variable in terms of a new
variable, $\beta=1-4\rho^{2}$, and pull out its square root, thus
defining new functions $\Delta_{i}$, 
\[
X_{i}\left(\rho\right)=\sqrt{\beta}\Delta_{i}\left(\beta\right),\qquad
i=0,1,2,A,F,L,S,H.
\]
The expansion around $\beta=0$ was first carried out in \cite{Czarnecki:1998kt}.
Our purpose in this section is to repeat that calculation, extend
it to higher powers in $\beta$, and make sure that the results match
the expansion around the zero-mass point, $\rho=0$, presented in
Section \ref{sec:Expansion-From-Zero}. The one-loop correction in
the $\beta$ expansion reads

\begin{eqnarray*}
\Delta_{1} & = & \frac{27}{8}\ln2-3+\beta\left(\frac{25}{8}\ln2+\frac{1}{2}\ln\beta-\frac{95}{48}\right)\\
 &  & +\beta^{2}\left(\frac{28}{15}\ln2+\frac{7}{15}\ln\beta-\frac{13483}{7200}\right)+\beta^{3}\left(\frac{44}{35}\ln2+\frac{11}{35}\ln\beta-\frac{143263}{117600}\right).\end{eqnarray*}
In \cite{Czarnecki:1998kt} the strong coupling constant was normalized
at the geometrical mean of the quark masses, $\alpha_{s}(\sqrt{m_{b}m_{c}})$,
while here we use $\alpha_{s}(m_{b})$, in order to be able to match
with the expansion around $\rho=0$. Also, in \cite{Czarnecki:1998kt},
the $c$-quark and $b$-quark loop contributions were added together
and denoted $\Delta_{H}$, while here we separate them. The $b$-quark
loop contribution is denoted by $\Delta_{H}$ and the $c$-quark by
$\Delta_{S}$. For reference, the $\Delta_{S}$ and $\Delta_{H}$
terms are given in Eqs.~(\ref{eq:DH},~\ref{eq:DS}) up to order
$\beta^{2}$ (both normalized with $\alpha_{s}\left(m_b\right)$),

\begin{eqnarray}
\Delta_{H} & = & \frac{509}{48}+\frac{999}{32}R_{2}+\frac{87}{16}\ln2+\frac{337}{64}\ln^{2}2+\frac{75\pi^{2}}{128}\label{eq:DH}\\
 &  & +\beta\left(\frac{7937}{864}+\frac{1449}{32}R_{2}+\frac{275}{144}\ln2+\frac{767}{64}\ln^{2}2+\frac{655\pi^{2}}{384}\right)\nonumber \\
 &  & +\beta^{2}\left(\frac{610309}{51840}+\frac{204969}{2560}R_{2}+\frac{59519}{17280}\ln2+\frac{973327}{46080}\ln^{2}2+\frac{317957\pi^{2}}{92160}\right),\nonumber \\
\Delta_{S} & = & \frac{361}{96}-\frac{621}{256}R_{2}-\frac{25}{64}\ln2-\frac{531}{512}\ln^{2}2-\frac{1445\pi^{2}}{3072}\label{eq:DS}\\
 &  & +\beta\left(\frac{433\pi^{2}}{3072}-\frac{757}{864}-\frac{207}{256}R_{2}-\frac{91}{576}\ln2-\frac{1}{3}\ln2\ln\beta-\frac{2579}{1536}\ln^{2}2\right)\nonumber \\
 &  & 
 +\beta^{2}\left[  \frac{287639}{414720}-\frac{3243}{20480}R_{2}+\frac{1120967}{691200}\ln2
 -\frac{85913}{73728}\ln^{2}2-\frac{51907\pi^{2}}{442368}
-\left(\frac{1}{6}+\frac{14}{45}\ln2\right)
\ln\beta \right],\nonumber \end{eqnarray}
where $R_2$ is obtained from \cite{Czarnecki:1997cf} and has a numerical value of $R_2 \approx -0.72964$.

While these changes were carried out, an error was noticed in the
charge renormalization used in \cite{Czarnecki:1998kt}. In that
paper, $\alpha_{s}$ was normalized at $\sqrt{m_{b}m_{c}}$. The error
consisted in using five quark flavors to run $\alpha_{s}$ down to
that scale, instead of excluding the $b$-quark in the range between
$m_{b}$ and $\sqrt{m_{b}m_{c}}$. This error originates in \cite{Czarnecki:1996gu}.
We have corrected for this in Eq.~(\ref{eq:DH}) and Table~\ref{tab:beta}. 

\begin{figure*}
\centering\includegraphics[width=0.25\textwidth]{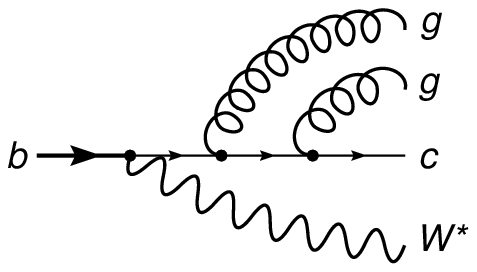}    
  \includegraphics[width=0.2\textwidth]{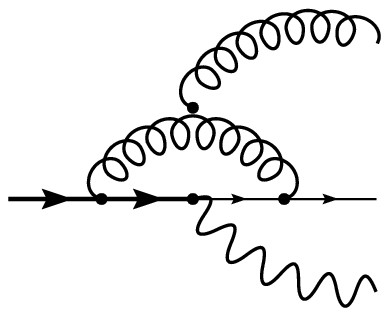}       \includegraphics[width=0.18\textwidth]{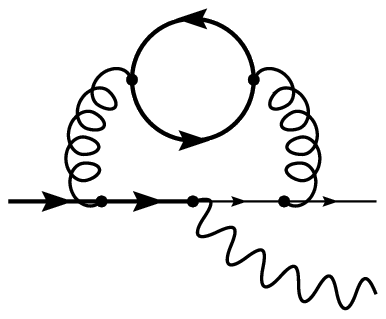}

\caption{\label{fig:PhaseSpace}A sample of the diagrams needed to calculate
the expansion presented in \cite{Czarnecki:1998kt} and updated here.}

\end{figure*}

To have proper matching between the expansion in \cite{Czarnecki:1998kt}
and the expansion presented here, we also found that the old expansion
needed more terms than could be obtained with the available computing
resources in 1998. We have updated the expansion to include analytical terms
up to $\beta^{8}$ as compared to $\beta^{4}$ previously. To carry
out this calculation, we used the same methods as the authors of \cite{Czarnecki:1998kt}.
Instead of calculating the corrections using self-energy diagrams
and the optical theorem, we calculated each second-order decay diagram
seperately. This required the calculation of 73 diagrams with zero,
one or two loops and up to four-particle phase space integrations,
Fig.~\ref{fig:PhaseSpace}.

In this expansion, the loops have been integrated using the same methods
described earlier. This lead to the calculation of 14 regions with
only one having an eikonal propagator \cite{Czarnecki:1996nr}. The
numerical coefficients of all terms that have been calculated here
are shown in Table~\ref{tab:beta}.

\begin{table*}
\begin{tabular}{c>{\centering}p{1.1cm}>{\centering}p{1.1cm}>{\centering}p{1.2cm}>{\centering}p{1.2cm}>{\centering}p{1.1cm}>{\centering}p{1.2cm}>{\centering}p{1.2cm}>{\centering}p{1.2cm}>{\centering}p{1.2cm}>{\centering}p{1.2cm}>{\centering}p{1.2cm}>{\centering}p{1.2cm}>{\centering}p{1.2cm}}
\hline 
\noalign{\vskip0.5mm}
 & $\beta^{0}$ & $\beta^{1}$ & $\beta^{1}\ln\beta$ & $\beta^{1}\ln^{2}\beta$ & $\beta^{2}$ & $\beta^{2}\ln\beta$ & $\beta^{2}\ln^{2}\beta$ & $\beta^{3}$ & $\beta^{3}\ln\beta$ & $\beta^{3}\ln^{2}\beta$ & $\beta^{4}$ & $\beta^{4}\ln\beta$ & $\beta^{4}\ln^{2}\beta$\tabularnewline
\hline
\hline 
\noalign{\vskip0.5mm}
$\Delta_{A}$ & $-1.849$ & 0.420 & 2.421 & $-0.458$ & $-1.960$ & 2.109 & $-0.428$ & $-1.411$ & 1.702 & $-0.288$ & $-0.730$ & 1.311 & $-0.218$\tabularnewline
\noalign{\vskip0.5mm}
$\Delta_{F}$ & 1.762 & $-0.854$ & $-0.440$ & - & 0.015 & $-0.140$ & 0.167 & 0.208 & $-0.650$ & 0.256 & 0.442 & $-0.642$ & 0.270\tabularnewline
\noalign{\vskip0.5mm}
$\Delta_{L}$ & 0.419 & 0.086 & $-0.982$ & 0.167 & 0.576 & $-0.804$ & 0.156 & 0.568 & $-0.712$ & 0.105 & 0.257 & $-0.512$ & 0.079\tabularnewline
\noalign{\vskip0.5mm}
$\Delta_{S}$ & 0.118 & 0.189 & $-0.231$ & - & 0.215 & $-0.382 $ & - & 0.323 & $-0.384$ & - & 0.257 & $-0.348$ & -\tabularnewline
\noalign{\vskip0.5mm}
$\Delta_{H}$ & $-0.087$ & 0.072 & - & - & $-0.045$ & - & - & $-0.002$ & - & - & $-5\times10^{-4}$ & - & -\tabularnewline
\hline
\end{tabular}\smallskip{}

\begin{tabular}{c>{\centering}p{1.2cm}>{\centering}p{1.2cm}>{\centering}p{1.2cm}>{\centering}p{1.2cm}>{\centering}p{1.2cm}>{\centering}p{1.2cm}>{\centering}p{1.2cm}>{\centering}p{1.2cm}>{\centering}p{1.2cm}>{\centering}p{1.2cm}>{\centering}p{1.2cm}>{\centering}p{1.2cm}}
\hline 
\noalign{\vskip0.5mm}
 & $\beta^{5}$ & $\beta^{5}\ln\beta$ & $\beta^{5}\ln^{2}\beta$ & $\beta^{6}$ & $\beta^{6}\ln\beta$ & $\beta^{6}\ln^{2}\beta$ & $\beta^{7}$ & $\beta^{7}\ln\beta$ & $\beta^{7}\ln^{2}\beta$ & $\beta^{8}$ & $\beta^{8}\ln\beta$ & $\beta^{8}\ln^{2}\beta$\tabularnewline
\hline
\hline 
\noalign{\vskip0.5mm}
$\Delta_{A}$ & $-0.429$ & 1.032 & $-0.176$ & $-0.306$ & 0.883 & $-0.147$ & $-0.223$ & 0.766 & 0.127 & $-0.174$ & 0.682 & $-0.111$\tabularnewline
\noalign{\vskip0.5mm}
$\Delta_{F}$ & 0.298 & $-0.670$ & 0.264 & 0.189 & $-0.494$ & 0.254 & 0.177 & $-0.493$ & 0.242 & 0.122 & $-0.410$ & 0.230\tabularnewline
\noalign{\vskip0.5mm}
$\Delta_{L}$ & 0.159 & $-0.413$ & 0.064 & 0.110 & $-0.349$ & 0.054 & 0.081 & $-0.304$ & 0.0462 & 0.062 & $-0.269$ & 0.041\tabularnewline
\noalign{\vskip0.5mm}
$\Delta_{S}$ & 0.191 & $-0.314$ & - & 0.147 & $-0.285$ & - & 0.117 & $-0.261$ & - & 0.096 & $-0.241$ & -\tabularnewline
\noalign{\vskip0.5mm}
$\Delta_{H}$ & $-1\times10^{-4}$ & - & - & $-5\times10^{-5}$ & - & - & $-2\times10^{-5}$ & - & - & $-1\times10^{-5}$ & - & -\tabularnewline
\hline
\end{tabular}

\caption{\label{tab:beta}Numerical coefficients to all orders calculated for
the updated expansion from the zero-recoil limit. The values have
been calculated using $\alpha_{s}(m_{b})$.}

\end{table*}

\begin{figure}[!bh]
\centering
\includegraphics[width=8cm]{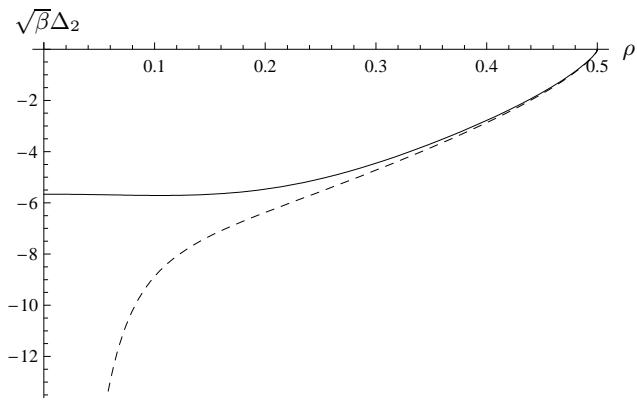}
\put(-225,140){$\sqrt{\beta}\Delta_2$}
\put(5,130){$\rho$}

\caption{\label{fig:Czar}Second order corrections expanded from the zero recoil
line. The dashed line shows the expansion up to $\mathcal{O}(\beta^{4})$
while the solid line shows the expansion up to $\mathcal{O}(\beta^{8})$.  For the purpose of comparing with \cite{Czarnecki:1998kt}, this plot is made assuming $\alpha_s(\sqrt{m_b m_c})$ is used in the NLO correction.}

\end{figure}

Fig.~\ref{fig:Czar} shows how the updated expansion differs from
the previous one and clearly displays the need for the higher order
terms, at and below the physical value $\rho\sim0.3$. 

In an attempt to account for the higher order terms, the authors of
\cite{Czarnecki:1998kt} added a term to approximate the remainder
of the series equal to the product of highest order term and $\frac{\beta}{2(1-\beta)}$.
This also gave an estimate of the error in their calculation. For
a value of $\rho=0.3$ ($\beta=0.64$) they found,
\begin{equation}
\sqrt{\beta}
\Delta_{2}=-4.72(14).\end{equation}
 With the extra terms we have calculated here and the corrections
to the charge renormalization, this changes to,\begin{equation}
\sqrt{\beta}\Delta_{2}=-4.45(1),\end{equation}
where we have used the same method of estimating the error. With an
error of $\approx0.2\%$, we have sufficient accuracy for computing
the full decay width $\Gamma(b\to c\ell\nu_{\ell})$ in the next section.

\begin{figure}[!b]
\includegraphics[width=0.45\textwidth]{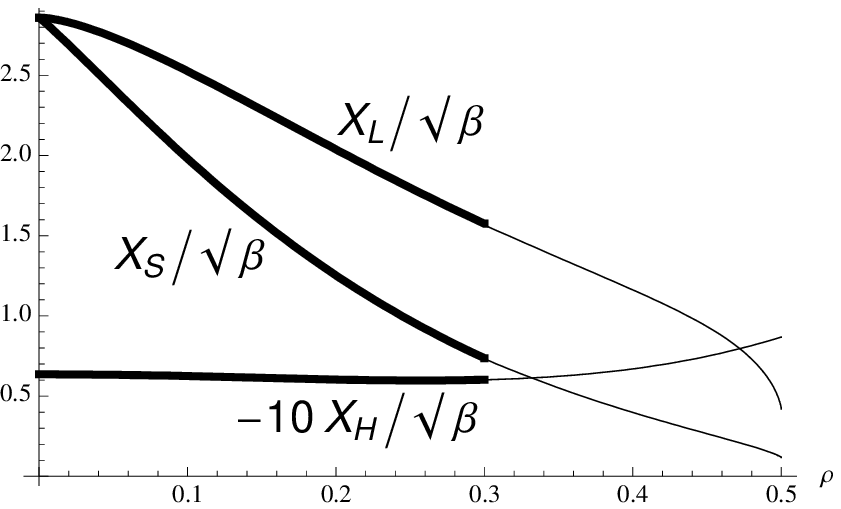}

\begin{minipage}[t][1\totalheight]{0.45\textwidth}%
Quark loop contributions%
\end{minipage}

\includegraphics[width=0.45\textwidth]{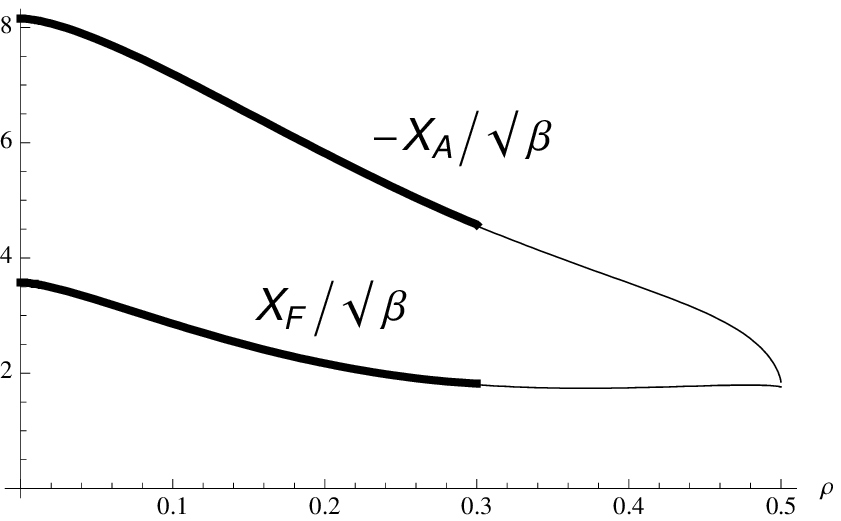}

\begin{minipage}[t][1\totalheight]{0.45\textwidth}%
Abelian and Non-Abelian contributions.%
\end{minipage}

\includegraphics[width=0.45\textwidth]{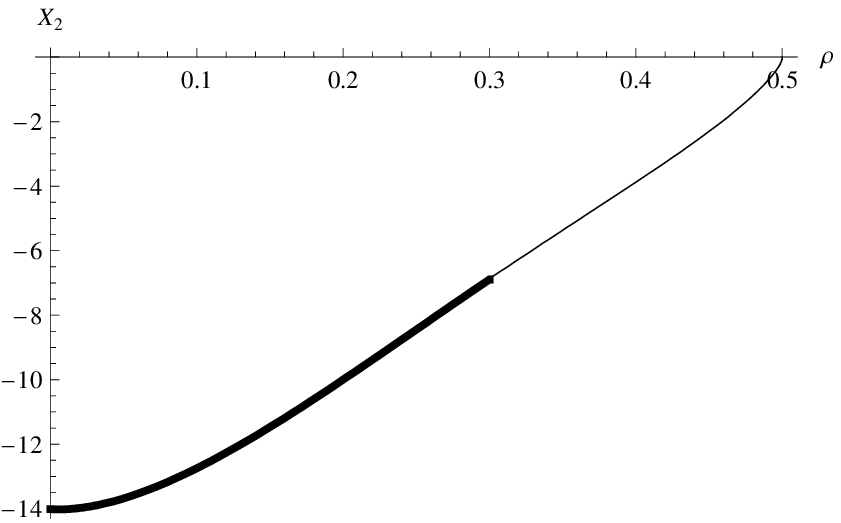}

\begin{minipage}[t][1\totalheight]{0.45\textwidth}%
Total $\mathcal{O}(\alpha_{s}^{2})$ correction.%
\end{minipage}

\caption{\label{fig:contrib}The matching between the different colour contributions
and total $X_{2}$ contribution. The thick line corresponds to the
expansions presented here and the thin line corresponds to the updated
expansion from the zero recoil line.}

\end{figure}

Comparing the two expansions, around $\rho=0$ and around $\rho=\frac{1}{2}$,
as shown in Fig.~\ref{fig:contrib}, one can now see that all of
the different colour contributions and thus the full $\alpha_{s}^{2}$
corrections agree very well.%

\section{Estimate of the full correction to the semileptonic decay rate}
\subsection{Notation}
So far in this paper we have been concerned with the decay of a $b$-quark into a $c$-quark and a virtual $W$-boson, at the intermediate recoil where the masses of $c$ and $W^*$ are equal.  We now want to use the results we have obtained, together with previously obtained values at zero- and maximum-recoil to fit the corrections to the decay $b\to c \ell \nu_{\ell}$.  We follow the notation of \cite{Czarnecki:1998kt},
\begin{eqnarray}
{d\Gamma(b\to c\ell \nu_{\ell}) \over d q^2}
&=& {G_F^2 m_b^3 |V_{cb}|^2 \over 96 \pi^3}
\left[ A_{\mathrm{Born}} + 
{\alpha_s(\sqrt{m_b m_c} ) \over \pi} C_F A_1
+ \left( {\alpha_s \over \pi} \right)^2 C_F A_2 
\right],
\nonumber \\
A_{\mathrm{Born}} &=&
\sqrt{(1-\rho^2-q^2)^2-4\rho^2 q^2}\,
\left[ (1-\rho^2)^2+ (1+\rho^2)q^2-2q^4\right].
\label{eq:diff}
\end{eqnarray}
In addition, we define the corrections for the integrated decay rate,
\begin{eqnarray}
\Gamma(b\to c\ell \nu_{\ell})
&=& {G_F^2 m_b^5 |V_{cb}|^2 \over 192 \pi^3}F(\rho)
\left[1 + 
{\alpha_s(\sqrt{m_b m_c} ) \over \pi} B_1
+ \left( {\alpha_s \over \pi} \right)^2  B_2 
\right],
\nonumber \\
F(\rho) &\equiv & 1 - 8\rho^2 - 24\rho^4\ln\rho + 8\rho^6 -\rho^8.
\end{eqnarray}
As we have already mentioned in the Introduction, the  NNLO
corrections $A_2$  and $B_2$ can be divided into the BLM and the non-BLM parts,
\begin{eqnarray}
A_2 &=& 
T_R(N_L A_L + N_S A_S + N_H A_H) + C_F A_F + C_A A_A
\nonumber \\
&\equiv &A^{\mathrm{BLM}} + A^{\mathrm{nBLM}},
\nonumber \\
A^{\mathrm{BLM}} & \equiv & A_L \left[ T_R (N_L + N_S) -{11\over
    4}C_A\right],
\end{eqnarray}
and similarly for the integrated corrections $B$.
All the functions $A$ in the Eq.~(\ref{eq:diff})  depend on two variables: the quark mass ratio $\rho$ and the invariant mass of the leptons $\sqrt{q^2}$.  The full dependence on these variables is not yet known.  The expansions such as described in this paper and earlier studies determine $A$'s along the sides and the bisector  of the triangle shown in Fig.~\ref{fig:Triangle}.  Of particular interest are their values along the vertical line corresponding to the physical value of $\rho \simeq 0.3$,  describing the differential decay rate $d\Gamma(b\to c\ell\nu_{\ell})/dq^{2}$.
Ref. \cite{Czarnecki:1998kt} used the three known points along that
line to fit a polynomial
and integrate Eq.~(\ref{eq:Integral}) over $q^{2}$, thus providing
an estimate of the second order non-BLM corrections to the full semi-leptonic decay width,
\begin{equation}
B_{\mathrm{fit}}^{\mathrm{nBLM}}=0.9(3).
\label{eq:CM}\end{equation}
This numerical value is quoted from \cite{Melnikov:2008qs}, where
the author fixed another mistake in \cite{Czarnecki:1998kt}, related
to using the maximum-recoil result. Recently, however, two calculations
of the full decay width $\Gamma(b\to c\ell\nu_{\ell})$ \cite{Melnikov:2008qs,Pak:2008qt}
gave
\begin{equation}
B^{\mathrm{nBLM}}=1.73(4).
\label{eq:AK}\end{equation}
This section is devoted to clarifying the discrepancy between these
results.

For comparison purposes we use $\alpha_{s}(\sqrt{m_{b}m_{c}}),\: N_{f}=4,\textrm{ and }\rho=0.3$
as in \cite{Czarnecki:1998kt}, where $N_{f}$ refers to the number
of light quarks used to calculate the BLM contributions. The authors
of \cite{Melnikov:2008qs,Pak:2008qt} use a different set of parameters, to which we will return in section \ref{sec:better}. 

\subsection{Effect of corrected coupling constant normalization}
\label{sec:normal}
A large part of the discrepancy between Eqs. (\ref{eq:CM},\ref{eq:AK}) 
is due to the incorrect charge renormalization, as discussed in Section \ref{sec:zero}. 
We have corrected this and recalculated the non-BLM contributions 
using the same fitting method described in \cite{Czarnecki:1998kt}. Analogously
to Eq.~(8) in \cite{Czarnecki:1998kt}, we introduce a new function of the lepton invariant mass $q^2$ at fixed quark-mass ratio $\rho$ (we use $\rho=0.3$).  It is denoted $\xi(q^2)$ and defined as  
\begin{eqnarray}
\xi(q^{2})&=&\frac{A_{2}(q^{2})-A_{2}^{\mathrm{BLM}}(q^{2})}{A_{\mathrm{Born}}(q^{2})},
\label{eq:Xi}
\end{eqnarray}
The three available values of $\xi(q^{2})$ at $q^{2}=0$, $m_{c}^{2}$,
and $q_{\mathrm{max}}^{2}=(m_{b}-m_{c})^{2}$ are 
\begin{eqnarray}
\xi(0)=1.26, & \xi(m_{c}^{2})=1.27, & \xi(q_{\mathrm{max}}^{2})=0.19.
\label{eq:Values}\end{eqnarray}
These values have been obtained using results from \cite{Pak:2006xf},
\cite{Czarnecki:1998kt}, and \cite{Czarnecki:1996gu}, with the
$b$-quark charge renormalization terms from \cite{Czarnecki:1996gu,Czarnecki:1998kt}
corrected. 
Fitting these values to a function defined by, 
\begin{equation}
\xi(q^{2})=a_{1}q^{4}+a_{2}q^{2}+a_{3},
\end{equation}
we integrate over $q^{2}$ to find a value for the non-BLM corrections.
The values quoted in Eq.~(\ref{eq:Values}) are normalized to the Born rate, $A_{\mathrm{Born}}$,
so the integral needed is analogous to Eq.~(9) in \cite{Czarnecki:1998kt},
\begin{equation}
B_{\mathrm{fit}}^{\mathrm{nBLM}}= C_F
\frac
{\int_{0}^{q_{\mathrm{max}}^{2}}dq^{2}
A_{\mathrm{Born}}(q^{2})\xi(q^{2})}
{\int_{0}^{q_{\mathrm{max}}^{2}}dq^{2}A_{\mathrm{Born}}(q^{2})}.
\label{eq:integration}
\end{equation}
This integration, with the input from Eq.~(\ref{eq:Values}),
gives $B_{\mathrm{fit}}^{\mathrm{nBLM}}=1.4(2)$. 
This agrees  with 
Eq.~(\ref{eq:AK}) much better than the value given in Eq.~(\ref{eq:CM}). 
The error is estimated by performing the analogous fit of the BLM corrections
and comparing the result to the exact value  \cite{Luke:1994yc}.

\subsection{Effect of extending the expansion to higher orders in $\beta$}

In section \ref{sec:normal} we have merely corrected the renormalization in the old results.  In addition, using the results of the two expansions in the present paper, we can obtain a more accurate input along the intermediate recoil line.  Instead of the value 1.27 in Eq. (\ref{eq:Values}), this gives $\xi(m_c^2)=1.33$.  This is related to the shift induced by the higher-order terms, illustrated in Fig.~\ref{fig:Czar}.  There we see that the full correction is less negative than previously assumed, thus the difference with the BLM correction is more positive (larger).  Since the zero- and the intermediate-recoil points are close to each other, even a small shift of the value at one of them may be amplified in the integral of the fitted function.

After the integration in Eq.~(\ref{eq:integration}), 
this change leads to the new value  $B_{\mathrm{fit}}^{\mathrm{nBLM}}=1.5(2)$ which  now agrees with Eq.~(\ref{eq:AK}). 
The error estimated by comparing with the BLM case is 
about 12 per cent. By correcting the mistake in renormalization
and including more terms in the expansion from zero recoil, we have
brought the disagreement from about a factor of two down to $\approx10$ per cent, 
within the quoted error margins.

\subsection{A better fitting method}
\label{sec:better}
Further improvement is possible using
a better method of fitting the polynomial. 
In Eq.~(\ref{eq:Xi}), we normalized the points
to the tree level result $A_{\mathrm{Born}}$. We find that, if this normalization
is not done, i.e. instead we use,
\begin{equation}
\zeta(q^{2})=A_{2}(q^{2})-A_{2}^{\mathrm{BLM}}(q^{2}),\label{eq:Zeta}\end{equation}
the polynomial fit gives a much better estimate of the exact result.
With this adjustment of the fitting procedure, we end up with a final
non-BLM estimate of $B_{2}^{\mathrm{nBLM}}=1.76(4)$, a significant improvement.
Without knowing the shape of the $d\Gamma(b\to c\ell\nu_{\ell})/dq^{2}$
curve, we cannot say whether this is a numerical coincidence.
We have also performed this fitting for the $\mathcal{O}(\alpha_{s})$
corrections and BLM approximation. Both estimates give results that
are within $\approx3$ per cent of the exact known result and are more accurate
than using an analog of Eq.~(\ref{eq:Xi}) for the fit.

In the more recent papers \cite{Melnikov:2008qs,Pak:2008qt} the
authors use a different set of parameters for their calculations,
namely $\alpha_{s}(m_b),\: N_{f}=3,\textrm{ and }\rho=0.25$. For easy
comparison, we have also calculated the non-BLM corrections with this
set of parameters. Using Eq.~(\ref{eq:Zeta}) for the fitting procedure
and the expansion about $\rho=0$ presented here, we find $B_{\mathrm{fit}}^{\mathrm{nBLM}}=3.37(15)$,
as compared with the result of $B^{\mathrm{nBLM}}=3.40(7)$ from \cite{Melnikov:2008qs}.

\section{Summary}
To summarize: we have corrected an error in the strong coupling constant
normalization in the previous intermediate-recoil expansion. We have
extended that expansion to several higher orders in the parameter
$\beta$, describing the difference between $m_{c}$ and $m_{b}/2$.
We have confirmed the correctness of that expansion by constructing
a new one, also along the intermediate-recoil diagonal but around
its other end, corresponding to $m_{c}/m_{b}\to0$. This analysis
allowed us to re-evaluate the fit of the $d\Gamma(b\to c\ell\nu_{\ell})/dq^{2}$
curve based on the three kinematical points, and remove the disagreement
between the correction to the total rate $\Gamma(b\to c\ell\nu_{\ell})$
obtained from such a fit, and that obtained from the direct four-loop
calculations \cite{Melnikov:2008qs,Pak:2008qt}. With this result,
the full massive calculation of the $\mathcal{O}\left(\alpha_{s}^{2}\right)$
corrections to the semileptonic $b$-quark decay rate is confirmed. 

\begin{acknowledgments}
This work was supported by Science and Engineering Research Canada.
\end{acknowledgments}

\appendix
\section{Contributions From $b\to c\overline{c}c W^{*}$}
The expansions used to obtain the maximum recoil point for our polynomial
fit were calculated in \cite{Czarnecki:1997hc} and \cite{Pak:2006xf}.
The two expansions agree very well except for the region with $m_{c}<\frac{m_{b}}{3}$.
This discrepancy can be attributed to the omission of the amplitude
of $b\to c\overline{c}cW^{*}$ from the expansion in \cite{Czarnecki:1997hc}.
For completeness, we have calculated this contribution for both maximum
recoil and intermediate recoil, as this expansion was also not included
in \cite{Czarnecki:1998kt}. These expansions are calculated as threshold
expansions in terms of $\delta$ given by,\begin{equation}
m_{c}=\frac{m_{b}}{3(4)}(1-\delta),\end{equation}
 where the $3(4)$ indicates the factor used when calculating the
maximum recoil, $3$, or intermediate recoil, $4$, expansion. The
methods used for both expansions are discussed in \cite{Czarnecki:1998qc}.
This calculation relies on the ability to reduce the four particle phase
space integrals needed, into a product of two particle phase spaces.

For the maximum recoil case, the expansion has been calculated up
to $\delta^{14}$, with the first four terms given here, \begin{eqnarray}
\Gamma(b\to c\overline{c}cW^{*})_{m_{W^{*}}=0} & = & \frac{\Gamma_{0}\alpha_{s}^{2}\sqrt{3}\delta^{6}C_{F}T_{R}}{5\pi}\\
 &  & \left(1+\frac{83}{56}\delta+\frac{7}{64}\delta^{2}+\frac{55}{896}\delta^{3}+\frac{753}{896}\delta^{4}+\ldots\right).\nonumber \end{eqnarray}
 For intermediate recoil, the expansion has been calculated up to
$\delta^{9}\sqrt{\delta}$,\begin{eqnarray}
\Gamma(b\to c\overline{c}cW^{*})_{m_{W^{*}}=m_{c}} & = & \frac{3\Gamma_{0}\alpha_{s}^{2}\delta^{3}\sqrt{\delta}C_{F}T_{R}}{140\sqrt{2}\pi}\\
 &  & \left(1+\frac{535}{108}\delta-\frac{137045}{85536}\delta^{2}+\frac{175277863}{13343616}\delta^{3}+\ldots\right).\nonumber \end{eqnarray}

\end{document}